\documentclass{nature}
\usepackage[utf8]{inputenc}
\usepackage{tabulary,amsmath,amsfonts,amssymb}
\usepackage{url,multirow,cancel,tfrupee}
\makeatletter
\AtBeginDocument{\@ifpackageloaded{textcomp}{}{\usepackage{textcomp}}}
\makeatother
\usepackage{colortbl}
\usepackage{xcolor}
\usepackage{pifont}
\usepackage{graphicx} 
\usepackage{physics} 
\usepackage{float} 
\usepackage{epsfig,amssymb} 
\usepackage[10pt]{moresize} 
\usepackage{subfig} 
\usepackage{wrapfig}
\usepackage[nointegrals]{wasysym}
\makeatletter

\begin{document}

\title{Demonstration of Entanglement Purification and Swapping Protocol to Design Quantum Repeater in IBM Quantum Computer}
  
\author{Bikash K. Behera$^{1}$\thanks{E-mail: bkb13ms061@iiserkol.ac.in }{ },
              Swarnadeep Seth$^{1}$\thanks{E-mail: ss13ms054@iiserkol.ac.in}{ },
              Antariksha Das$^{1}$\thanks{E-mail: ad13ms055@iiserkol.ac.in} {} \&
              Prasanta K. Panigrahi$^{1}$\thanks{Corresponding author.}\ \thanks{E-mail: pprasanta@iiserkol.ac.in}
              }
\maketitle 

\begin{affiliations}
    \item
      Department of Physical Sciences\unskip, 
    Indian Institute of Science Education and Research Kolkata\unskip, Mohanpur\unskip, 741246\unskip, West Bengal\unskip, 
    India
\end{affiliations}
      
\begin{abstract}
Quantum communication is a secure way to transfer quantum information and to communicate with legitimate parties over distant places in a network. Although communication over a long distance has already been attained, technical problem arises due to unavoidable loss of information through the transmission channel. Quantum repeaters can extend the distance scale using entanglement swapping and purification scheme. Here we demonstrate the working of a quantum repeater by the above two processes. We use IBM's real quantum processor `ibmqx4' to create two pair of entangled qubits and design an equivalent quantum circuit which consequently swaps the entanglement between the two pairs. We then develop a novel purification protocol which enhances the degree of entanglement in a noisy channel that includes combined errors of bit-flip, phase-flip and phase-change error. We perform quantum state tomography to verify the entanglement swapping between the two pairs of qubits and working of the purification protocol. 
\end{abstract}
    
\section{Introduction}
Quantum communication \cite{DuaNNat2001,GisinNatPhoton2007,SongChinSciBull2012,OrieuxJOpt2016} is one of the secure ways to send unknown quantum states from one place to another and transfer secret messages among the parties. Photonic channels \cite{ZukowskiPRL1993,BriegelPRL1998,PengPRL2007,RosenbergPRL2007,SchmittPRL2007,ZhaoPRL2007,UrsinNphys2007} have been found a significant attraction for the physical implementation of quantum communication. The secret messages can be potentially transmitted through the photonic channel by using quantum cryptography \cite{EkertPRL1991,SimonEPJD2010,lct2014}, that plays a key role in building a quantum network \cite{ElliottNewJPhys2002,KimbleNat2008,TangNatSciRev2017}. The mechanism of quantum communication lies in generating entangled states \cite{MatsuokaPRA1993,AspelmeyerSci2003,ReschOptExpr2005,PengPRL2005,DasPRA2018} between distant parties, which is a difficult task to achieve practically \cite{Pirandola}. Using entangled channel, quantum teleportation protocol \cite{BennettPRL1993,BouwmeesterNat1997,RiebeNat2004,BarrettNat2004,UrsinNat2004} can be performed securely over a long distance. However, the degree of entanglement between distant parties decreases exponentially over a photonic channel even after using a purification scheme \cite{BennettPRL1991}. Hence, it becomes nearly impossible to keep intact the entangled state over a large scale distance.

Efficient long distance communication over the distances of the order 1000 km has remained an outstanding challenge due to loss errors in the communication channel. Quantum repeaters (QRs) have been proposed as promising candidates to overcome this problem. The purpose is to divide the whole distance into smaller segments with a length comparable to the attenuation length of the channel and establishing Quantum repeater stations \cite{DurPRA1999,LoockPRL2006,JiangPRA2009,FowlerPRL2010,MunroNatPhot2012} at each segment. This requires generation and purification of the entanglement for each segment and then transmission of the purified entanglement \cite{ChouScience2007} to the next segment through entanglement swapping \cite{BennettPRL1993,ZukowskiPRL1993,PanPRL1998,RiebeNATPhys2008}. The process of entanglement swapping and purification between two consecutive segments need to be repeated a large number of times until the entangled channel has been prepared with a high fidelity. To increase the fidelity of entangled state through the channel, quantum repeaters are introduced by Briegel, D\"{u}r, Cirac and Zoller (BDCZ) \cite{BriegelPRL1998}, which could in principle be used to preserve the entangled state with a fidelity close to unity. Other useful techniques such as heralded entanglement generation (HEG) \cite{BriegelPRL1998,SangouardRMP2011} or quantum error correction (QEC) \cite{JiangPRA2009,MunroNatPhot2010,FowlerPRL2010,MuralidharanPRL2014} have been applied to get rid of loss or operational errors \cite{MuralidharanSciRep2016} in the communication channel. Muralidharan \emph{et al.} \cite{MuralidharanSciRep2016} have listed a number of methods to overcome loss or operation errors and have classified three generations of quantum repeaters. Among all the methods, quantum error correction is a novel one as it can be used in all the generations of quantum repeaters as a purification protocol. Illustration of entanglement swapping and purification at each node fails to provide a complete demonstration of a quantum repeater protocol \cite{ChenzbPRA2007,PRA2008}. A quantum memory \cite{TrugenbergerPRL2001,gm2012} is in fact needed to store the entangled state of each repeater station, which is a great challenge to establish. However, Yuan \emph{et al.} \cite{YuanNature2008} have developed the scheme of a quantum memory while realizing the BDCZ quantum repeater, where, the process of entanglement swapping has been integrated with quantum memory for storage and retrieval of each segment state after the process of purification. 

IBM has developed prototypes of 5-, 16-, 20- and 50-qubit quantum computer \cite{quantumcomputing}, which has attracted the attention of a large number of researchers working in various sub-field of quantum computation and quantum information. A number of experiments have been tested and verified using 5-qubit and 16-qubit quantum computer \cite{HBP17,MajumderarXiv2017,SisodiaQIP2017, BKB6arXiv2017,WoottonQST2017,BertaNJP2016,DeffnerHel2017,HuffmanPRA2017,AlsinaPRA2016,GarciaarXiv2017,DasarXiv2017,BKB1QIP2017,YalcinkayaPRA2017,GhosharXiv2017,KandalaNAT2017,Solano2arXiv2017,SchuldEPL2017,SisodiaPLA2017}. Here, we use IBM's 5 qubit quantum processor `ibmqx4' to demonstrate the working of a quantum repeater. We entangle two pairs of superconducting qubits \cite{RJSScience2013} and design a quantum circuit which could in principle equivalently perform the main operations of a quantum repeater, i.e., entanglement swapping and entanglement purification. We introduce errors in the channel for generating unpurified entangled state and apply purification protocol on the erroneous channel that leads to the enhancement in the fidelity of the entangled state. It is to be mentioned that the entangled state is stored in the qubits of ibmqx4. The information about the entangled state can be retrieved by using an ancilla \cite{JRSIOP2010} representing another superconducting qubit. Hence, ibmqx4 can act as a quantum memory for storage and retrieval of information about the entangled state at each repeater node. In the transmission channel, if at every station, a quantum computer could be placed then that can act as a quantum repeater node. Therefore, quantum communication can be securely achieved by the help of quantum computers connected over the quantum internet \cite{ElliottNewJPhys2002,KimbleNat2008,TangNatSciRev2017} in a quantum networking environment. 
 
\section{Results} 

\begin{figure*}
\begin{center}
\epsfig{file=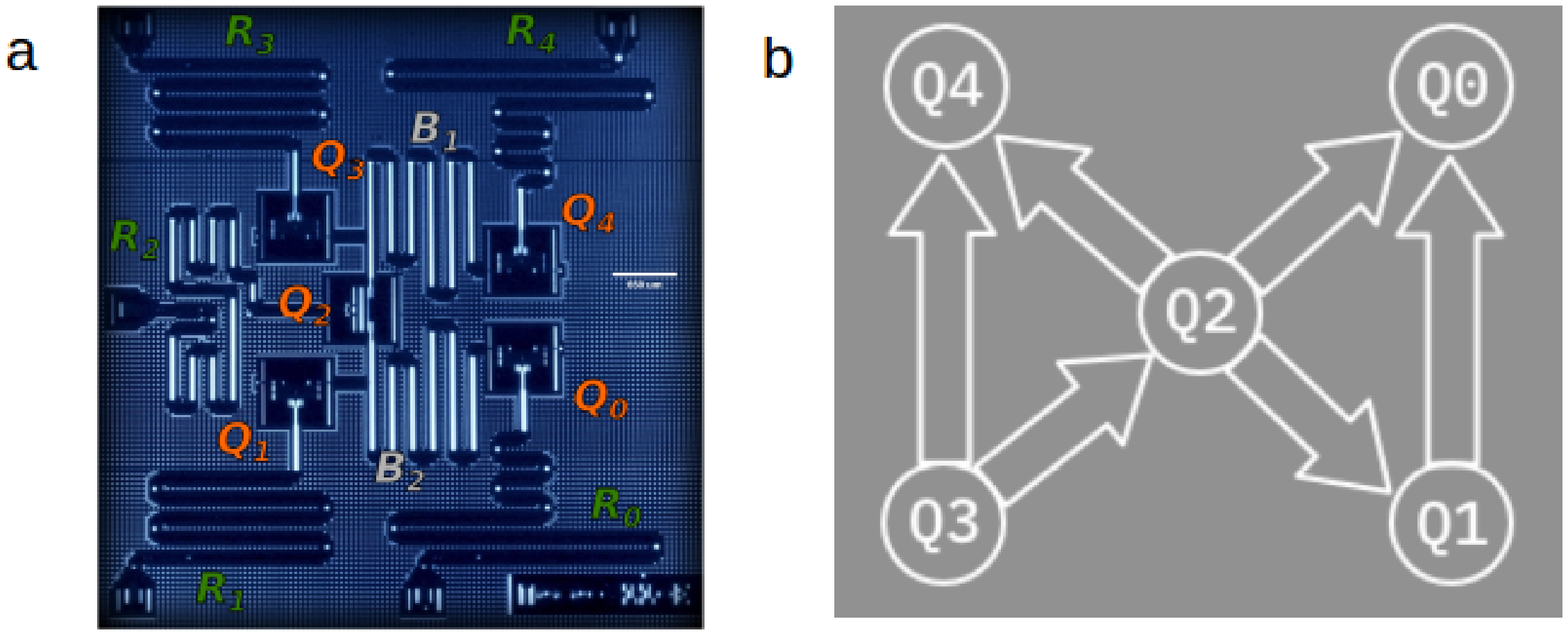,width=16cm}
\end{center}
\caption{The figure illustrates the chip layout of 5-qubit quantum processor ibmqx4. The chip is stored in a dilution refrigerator at temperature 0.021 K.  Here, all the 5 transmon qubits (charge qubits) are connected by two coplanar waveguide (CPW) resonators. The two CPWs couple Q2, Q3 and Q4 qubits with resonating frequency around 6.6 GHz and Q0, Q1 and Q2 qubits are coupled with 7.0 GHz frequency. Each qubit is controlled and readout by a particular CPW. (b) The coupling map for the CNOTS is represented as,$\{Q1 \rightarrow [Q0], Q2 \rightarrow [Q0, Q1, Q4], Q3 \rightarrow [Q2, Q4]\}$, where $a \rightarrow [b]$ means $a$ is the control qubit and $b$ is the target qubit for the implementation of CNOT gate. The gate and readout errors are of the order of $10^{-2}$ to $10^{-3}$.}
\label{fig1}
\end{figure*}

\textbf{Experimental Setup}: The experimental parameters of ibmqx4 chip are presented in Table \ref{tab1}, where $\omega^{R}_{i}$, $\omega_{i}$, $\delta_{i}$, $\chi$, $T_1$ and $T_2$ represent the resonance frequency, qubit frequency, anharmonicity, qubit-cavity coupling strength, relaxation time and coherence time respectively for the readout resonator. The connectivity and control of five superconducting qubits (Q0, Q1, Q2, Q3 and Q4) are depicted in Fig. \ref{fig1} (b). The single-qubit and two-qubit controls provided by the coplanar wave guides (CPWs) are shown by black and white lines respectively. The device is cooled in a dilution refrigerator at temperature 0.021 K. The qubits are coupled via two superconducting CPWs, one coupling Q2, Q3 and Q4 and another one coupling Q0, Q1, Q2 with resonator frequencies 6.6 Hz and 7.0 Hz respectively. The qubits are controlled and read out by individual CPWs. 
 
\begin{table}
\centering
\begin{tabular}{ c c c c c c c }
\hline
\hline
Qubits & $\omega^{R^{\bigstar}}_{i}/2\pi$ (GHz) & $\omega^{\dagger}_{i}/2\pi$ (GHz) & $\delta^{\ddagger}_{i}/2\pi$ (MHz) & $\chi^{\S}/2\pi$ (kHz) & $T^{||}_{1}$ ($\mu s$) & $T^{\perp}_{2}$ ($\mu s$)\\
\hline
\hline
Q0 & 6.52396 & 5.2461 & -330.1 & 410 & 35.2 & 38.1 \\
Q1 & 6.48078 & 5.3025 & -329.7 & 512 & 57.5 & 40.5 \\
Q2 & 6.43875 & 5.3025 & -329.7 & 408 & 36.6 &54.8 \\ 
Q3 & 6.58036 & 5.4317 & -327.9 & 434 & 43.0 & 42.1 \\
Q4 & 6.52698 & 5.1824 & -332.5 & 458 & 49.5 & 19.2\\
\hline
\hline
\end{tabular}
$\bigstar$ Resonance frequency, $\dagger$ Qubit frequency, $\ddagger$ Anharmonicity, $\S$ Qubit-cavity coupling strength, $||$ Relaxation time, $\perp$ Coherence time.
\caption{\textbf{The table shows the parameters of the device ibmqx4.}}
\label{tab1}
\end{table}

\begin{figure*}
\begin{center}
\epsfig{file=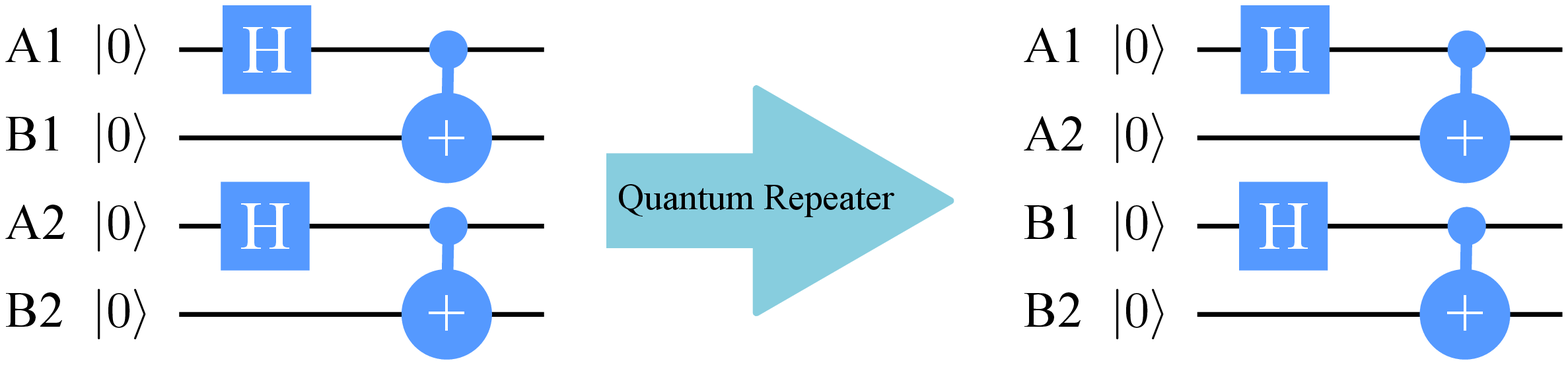,width=16cm}
\end{center}
\caption{The schematic diagram illustrates the performance of a Quantum Repeater (QR). The entanglement between the pair of qubits $A_1$-$B_1$ and $A_2$-$B_2$ is being swapped (using entanglement swapping protocol) to generate entanglement between $A_1$-$A_2$ and $B_1$-$B_2$ qubit pairs, which is an essential mechanism of a QR. A purification protocol is introduced to improve the fidelity of the entanglement between the above two pairs. These two mechanisms are applied to inherently represent the working of the QR.}
\label{fig2}
\end{figure*}

\begin{figure*}
   \begin{center}
   \epsfig{file=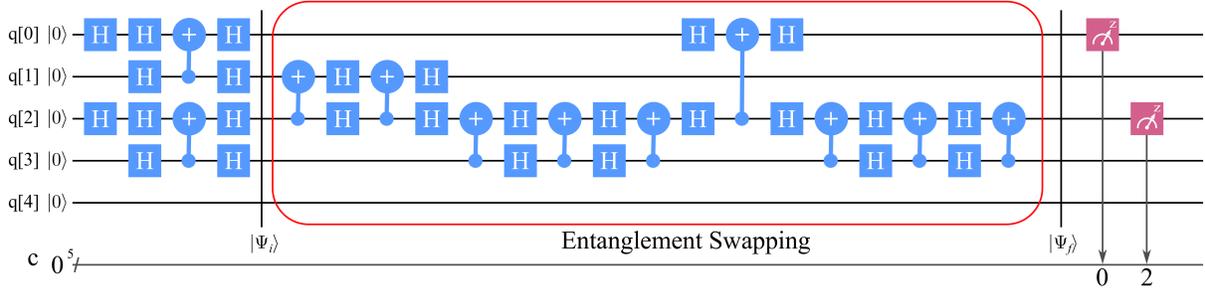,width=16cm}
   \end{center}
    \caption{\textbf{Entanglement swapping:} Initially, the qubits q[0], q[[1] and q[2], q[3] are each entangled by the Bell channel, $\frac{|00\rangle+|11\rangle}{\sqrt{2}}$. The pairs represent the entanglement between Alice and Charlie, and Bob and Charlie. After the entanglement swapping protocol, q[0], q[2] and q[1], q[3] are entangled, illustrating the entanglement between Alice and Bob, and Charlie's two qubits respectively. The experimental results showing the entanglement between the above parties are depicted in Figs. \ref{fig4} \& \ref{fig5}.}
    \label{fig3}
\end{figure*}

\textbf{Entanglement Swapping}: Entanglement swapping and purification of entanglement are the two main operations of a quantum repeater. We present the two schemes along with their experimental realization in the IBM quantum computer, ibmqx4. Entanglement swapping between two repeater stations is an essential condition for transferring information to distant places. This process can be understood by considering three parties, Alice, Bob and Charlie. We consider two entangling pairs of qubits, $A_1$-$B_1$, $A_2$-$B_2$, where $A_1$, $A_2$ stand for Alice and Bob's qubit respectively and $B_1$ and $B_2$ correspond to Charlie's qubit. Initially, Alice and Bob's qubits $A_1$ and $A_2$ were entangled with Charlie's qubits $B_1$ and $B_2$ respectively. After the swapping process, Alice and Bob's qubits, $A_1$ and $A_2$ get entangled, which is the key idea of entanglement swapping. In our experiment, we model the same scenario by means of superconducting qubits \cite{RJSScience2013} in IBM quantum experience interface \cite{quantumcomputing}. It is clearly seen from the Figs. \ref{fig2} \& \ref{fig3} that initially, $A_1$, $B_1$ and $A_2$, $B_2$ are entangled by the Bell channel, $\frac{|00\rangle+|11\rangle}{\sqrt{2}}$, meaning the entanglement between Alice and Charlie, and Bob and Charlie respectively. We design an equivalent quantum circuit that performs entanglement swapping between the above two pairs of qubits so that $A_1$,$A_2$ and $B_1$,$B_2$ get entangled. Now, Alice and Bob got entangled while Charlie's two qubits also got entangled. The above scheme is depicted in Fig.~\ref{fig3}. 

The process of entanglement swapping is presented as the following calculation. The initial state of the whole system is denoted as, 
\begin{equation}
\ket{\Psi_{i}}=\left(\frac{\ket{0_{A_1}0_{B_1}}+\ket{1_{A_1}1_{B_1}}}{\sqrt{2}} \right) \otimes \left( \frac{\ket{0_{A_2}0_{B_2}}+\ket{1_{A_2}1_{B_2}}}{\sqrt{2}}\right)
\end{equation}
After sequentially applying CNOT$_{3\rightarrow 2}$, CNOT$_{2\rightarrow 3}$ and CNOT$_{1\rightarrow 4}$, the final state is obtained as,
\begin{equation}
\ket{\Psi_{f}}=\left(\frac{\ket{0_{A_1}0_{A_2}}+\ket{1_{A_1}1_{A_2}}}{\sqrt{2}} \right) \otimes \left( \frac{\ket{0_{B_1}0_{B_2}}+\ket{1_{B_1}1_{B_2}}}{\sqrt{2}}\right)
\end{equation}
Here, CNOT$_{a \rightarrow b}$ is applied on the target qubit $b$, where $a$ acts as the control qubit. The quantum circuit for the above operation is shown in Fig.~\ref{fig3}. It is to be pointed that the two protocols given in Ref. \cite{BKB1QIP2017} have been used to design the quantum circuit in ibmqx4. 

We perform quantum state tomography to characterize the quantum states \cite{SisodiaPLA2017,Xin,Cramer} obtained in our experiment. This technique includes a comparison between theoretical and experimental density matrices. 

The theoretical density matrix of the initially prepared quantum state is given by,
\begin{equation}
\rho^{T}= \ket{\Psi}\bra{\Psi}    
\end{equation}
and the expression for the experimental density matrix for two qubit system is represented as,
\begin{equation}
\rho^{E}=\frac{1}{2^{2}}\sum_{i_{1},i_{2}=0}^{3}T_{i_{1}i_{2}}(\sigma_{i_{1}} \otimes \sigma_{i_{2}})
\label{eq22}
\end{equation}
where
\begin{equation}
T_{i_{1}i_{2}}=S_{i_{1}} \times S_{i_{2}}
\end{equation}
and the indices $i_{1}$ and $i_{2}$ can take values 0, 1, 2 and 3 corresponding to I, X, Y and Z Pauli matrices respectively. The Stokes parameters are described as, $S_{0}=P_{\ket{0}}+P_{\ket{1}}$,
$S_{1}=P_{\ket{0_{X}}}-P_{\ket{1_{X}}}$,
$S_{2}=P_{\ket{0_{Y}}}-P_{\ket{1_{Y}}}$,
$S_{3}=P_{\ket{0_{Z}}}-P_{\ket{1_{Z}}}$, where P represents the probability for the corresponding bases given in the subscript.

\begin{figure*}
   \begin{center}
   \epsfig{file=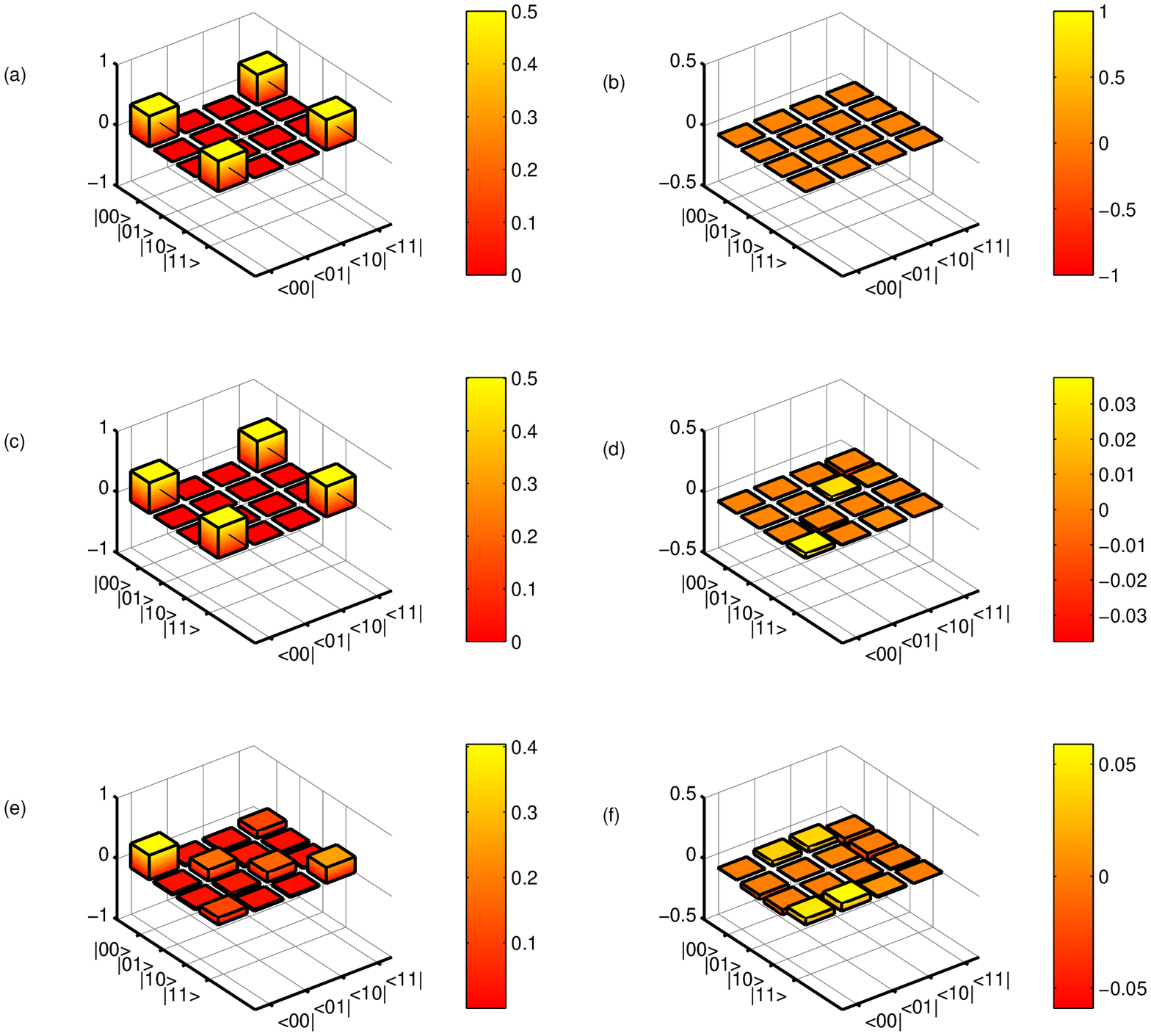,width=16cm}
   \end{center}
    \caption{The figure depicts both the real and imaginary parts of ideal, simulated and experimental density matrices for the $A_1$-$A_2$ entangled state. (a), (b): Ideal case; (c), (d): Simulated case; (e), (f): Experimental case. The entanglement between $A_1$-$A_2$ confirms the entanglement between Alice and Bob. The results are obtained after applying the entanglement swapping protocol shown in Fig. \ref{fig3} and measuring the qubits q[0] and q[2].}
    \label{fig4}
\end{figure*}

\begin{figure*}
   \begin{center}
   \epsfig{file=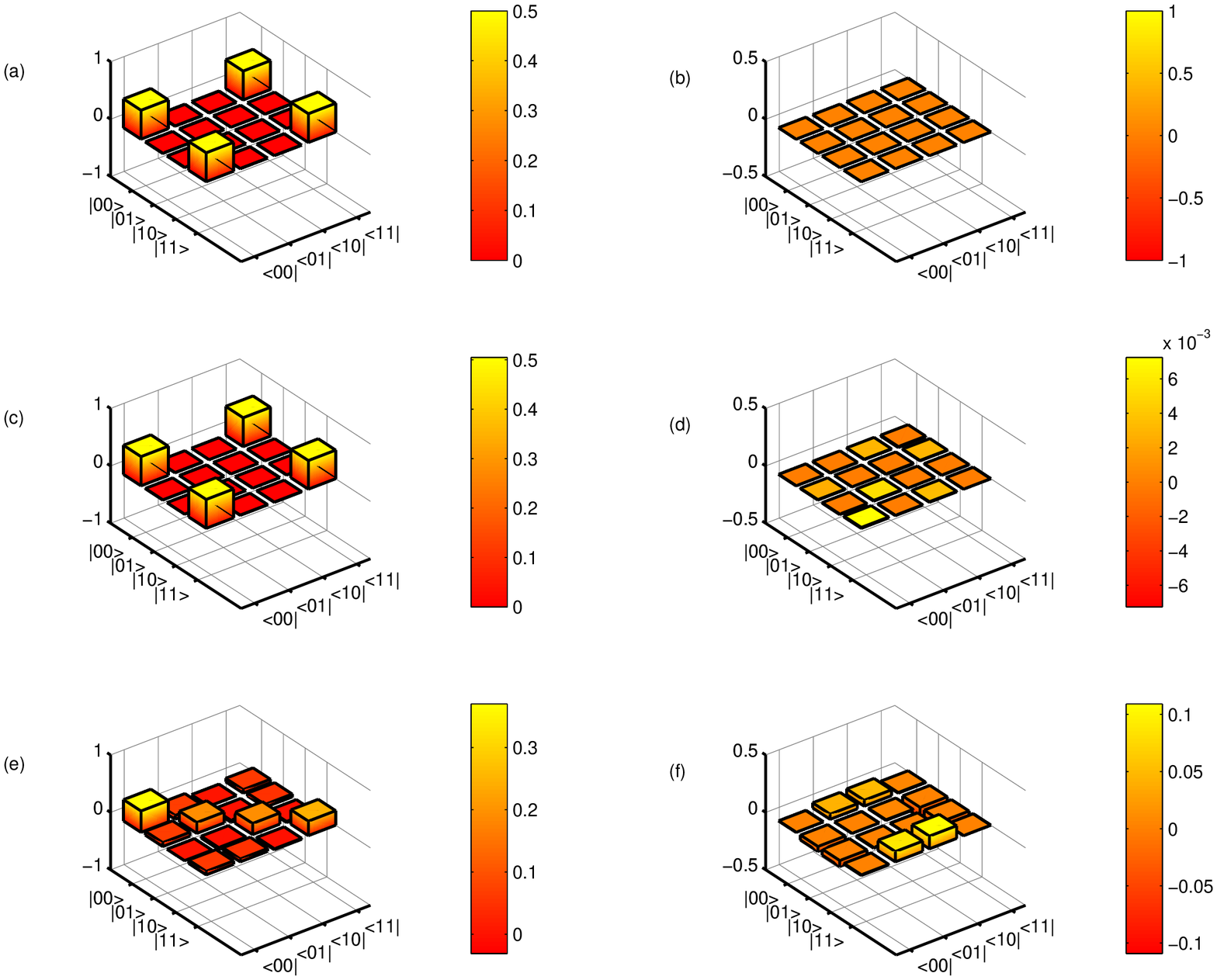,width=16cm}
   \end{center}
    \caption{ The figure depicts both the real and imaginary parts of ideal and experimental density matrices for the $B_1$-$B_2$ entangled state. (a,b): Ideal case. (c,d): Simulated case. (e,f): Experimental case. The entanglement between $B_1$-$B_2$ confirms the entanglement between Alice and Bob. The results are obtained after applying the entanglement swapping protocol shown in Fig. \ref{fig3} and measuring the qubits q[1] and q[3].}
    \label{fig5}
\end{figure*}

The fidelity \cite{Wimberger} between ideal and prepared arbitrary states of qubits $A$ and $B$ is calculated from,
\begin{equation}
\begin{split}
F(\rho^{T},\rho^{E})&=Tr\left(\sqrt{\sqrt{\rho^{T}}\rho^{E}\sqrt{\rho^{T}}}\right)\\
&=Tr\left(\sqrt{\ket{\Psi}\bra{\Psi}\rho^{E}\ket{\Psi}\bra{\Psi}}\right)
\end{split}
\end{equation}
The comparison among the density matrices for the above two cases are illustrated in Figs.~\ref{fig3} and \ref{fig4}. The fidelity of this experiment is calculated to be $F_{A_1A_2}=$0.8086 and $F_{B_1B_2}=$0.7840.

\textbf{Purification Protocol}: For different generation of quantum repeaters \cite{MuralidharanSciRep2016}, quantum error correction codes are used as entanglement purification protocol. In this scheme, we introduce noise in the channel to replicate the loss and operational errors which happen in a realistic situation. All types of errors i.e., combined error of bit-flip, phase-flip and phase-change errors are taken into account to make the channel noisy. Three single qubit gates X, U1($\pi$) and U1(0.125) are used for introducing bit-flip, phase-flip and phase-change errors respectively. The purification protocol is designed such that it can correct all types of errors in the noisy channel. The scheme uses Hadamard gates (H), phase gate (U1(-0.125)) and CNOT gates and more importantly a single ancilla quibit to rectify the introduced errors. The initial state is given as follows,
\begin{equation}
|\Psi_{Initial} \rangle=\frac{|00\rangle+|11\rangle}{\sqrt{2}}
\end{equation}
After introducing all types of errors, the unpurified state becomes
\begin{equation}
|\Psi_{Unpurified}\rangle=\frac{|01\rangle-e^{i\phi}|01\rangle}{\sqrt{2}}
\end{equation}
Applying the bit-flip, phase-flip and phase-change error correction codes, the purified state is
\begin{align}
&\frac{|00\rangle-e^{i\phi}|11\rangle}{\sqrt{2}} \quad \quad \textrm{(bit-flip error correction)} \nonumber \\
&\frac{|00\rangle+e^{i\phi}|11\rangle}{\sqrt{2}} \quad \quad \textrm{(phase-flip error correction)}\nonumber \\
|\Psi_{Purified}\rangle=&\frac{|00\rangle+|11\rangle}{\sqrt{2}} \quad \quad \textrm{(phase-change error correction)}
\end{align}
The fidelity before purification was $F_{BP}$=0.2891 and after purification it improves to $F_{AP}$=0.8456. The experimental result (See Fig.~\ref{fig7}) confirms that there is a high degree of entanglement in the channel after the purification.

\begin{figure*}
   \begin{center}
   \epsfig{file=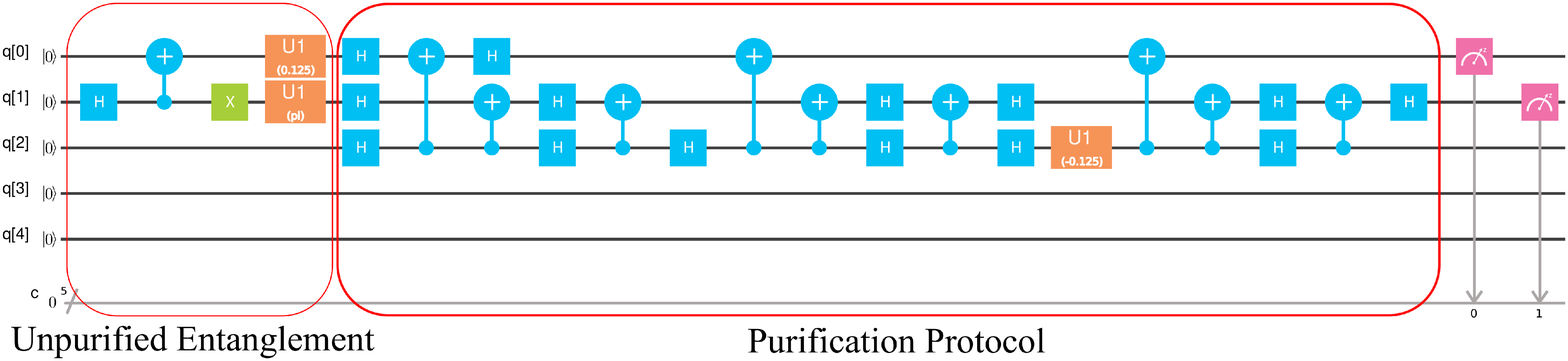,width=16cm}
   \end{center}
    \caption{\textbf{Purification Protocol:} The entanglement channel is unpurified by introducing noise, which includes all types of combined errors of bit-flip, phase-flip, and phase-change error. Three gates X, U1($\pi$) and U1(0.125) are used to introduce bit-flip, phase-flip and phase-change errors respectively. All the errors are then removed by the application of purification protocol, which uses only one ancilla qubit q[2]. The experimental results are obtained by measuring the first two qubits q[0] and q[1], which clearly shows the enhancement in the degree of entanglement of the channel. The results are depicted in Fig. \ref{fig7}. This protocol can be applied to any of the entangled channel between Alice and Bob and Charlie's qubits.}
    \label{fig6}
\end{figure*}

\begin{figure*}
   \begin{center}
   \epsfig{file=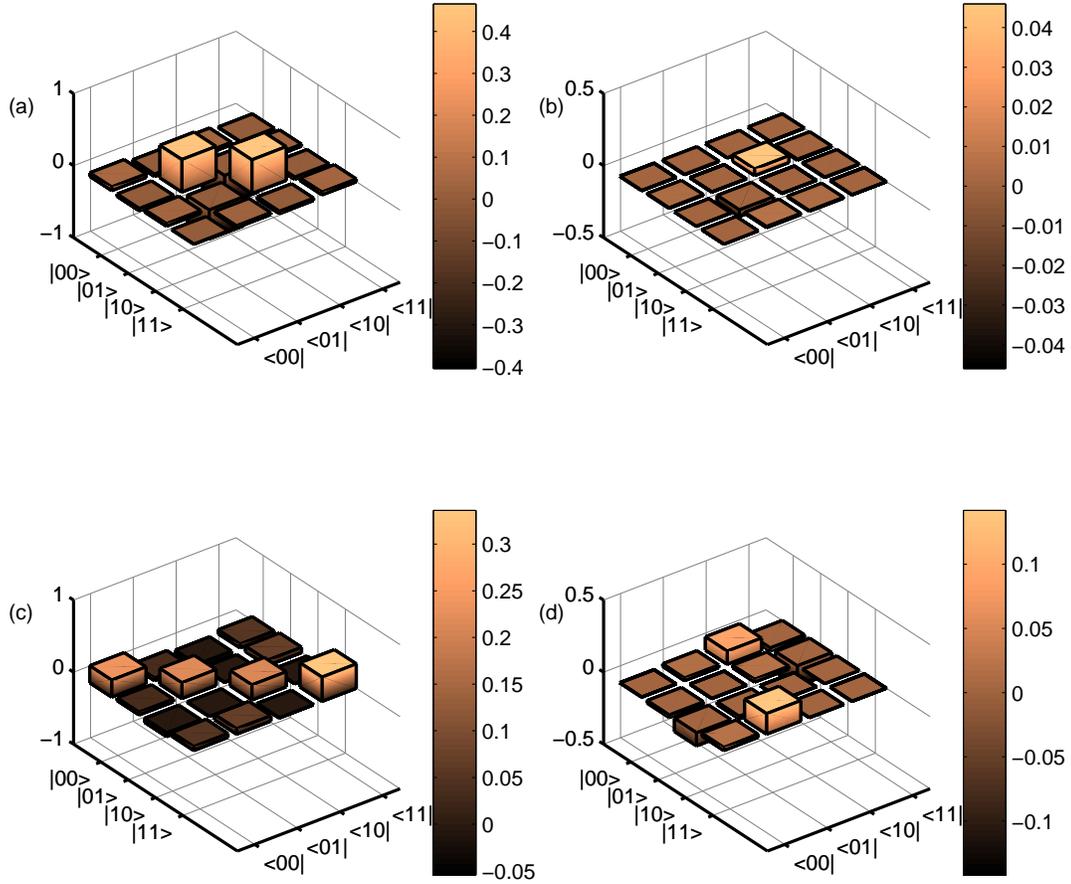,width=16cm}
   \end{center}
    \caption{ The figure depicts both the real and imaginary parts of unpurified and purified density matrices for the entangled state $\frac{|00\rangle+|11\rangle}{\sqrt{2}}$. (a,b): Before Purification. (c,d): After Purification. The fidelity of the unpurified state was 0.2891 and after purification fidelity improves to 0.8456, which clearly demonstrates the successful implementation of the purification protocol. }
    \label{fig7}
\end{figure*}

\section{Conclusions}
To conclude, we have explicated here the efficient processes of entanglement swapping \cite{BennettPRL1993,ZukowskiPRL1993,PanPRL1998,RiebeNATPhys2008} and entanglement purification scheme \cite{JiangPRA2009,MunroNatPhot2010,FowlerPRL2010,MuralidharanPRL2014}, two integral components of a Quantum Repeater, using superconducting qubits in IBM quantum computer (ibmqx4). Firstly, we have designed a quantum circuit which essentially swaps the entanglement between Alice, Bob and Charlie where Alice-Charlie and Bob-Charlie were initially entangled. After the entanglement swapping, Alice and Bob get entangled along with generating entanglement between Charlie's qubits. The entanglement between the parties Alice-Bob and charlie's two qubits are experimentally observed with fidelities $F_{A_1A_2}=$0.8086, $F_{B_1B_2}=$0.7840 respectively. During the process, the degree of entanglement substantially decreases due to the loss and operational errors in the channel. Hence, we have developed a robust purification protocol which can correct all types of operational errors \cite{DasarXiv2017} i.e., bit-flip, phase-flip and phase-change error that can occur in a noisy channel. the successful application of the purification protocol enhances the fidelity of entanglement from 0.2891 to 0.8456. Furthermore, establishing quantum memory to store the entangled state is a challenging task \cite{BriegelPRL1998}. Here, IBM quantum computer serves the purpose of a quantum memory by storing and retrieving the entangled states at each repeater node. Hence, we have successfully demonstrated a scheme of a Quantum Repeater for secure communication over a long distance in a quantum network.

\section*{References}
\bibliographystyle{naturemag}

\bibliography{\jobname}

\section*{Acknowledgements} B.K.B, S.S. and A.D. are financially supported by DST Inspire Fellowship. We are extremely grateful to IBM team and IBM Quantum Experience project. The discussions and opinions developed in this paper are only those of the authors and do not reflect the opinions of IBM or IBM QE team. 

\section*{Author contributions}
All authors have contributed in designing and developing the quantum circuits, performing experiment and writing the manuscript.  

\section*{Competing interests}
The authors declare no competing financial interests. 

\end{document}